# Spin injection characteristics of Py/graphene/Pt by gigahertz and terahertz magnetization dynamics driven by femtosecond laser pulse

H. Idzuchi[1-3*], S. Iihama[4,5#], M. Shimura[6], A. Kumatani[1,2,6,7], S. Mizukami[1,2,5], Y. P. Chen[3,8,9,1,2,5]

[1] *WPI Advanced Institute for Materials Research (AIMR), Tohoku University Sendai 980-8577, Japan*

[2] *Center for Science and Innovation in Spintronics (CSIS), Tohoku University Sendai 980-8577, Japan*

[3] *Purdue Quantum Science and Engineering Institute and Department of Physics and Astronomy, Purdue University, West Lafayette, Indiana 47907, USA*

[4] *Frontier Research Institute for Interdisciplinary Sciences (FRIS), Tohoku University, Sendai 980-8578, Japan*

[5] *Center for Spintronics Research Network (CSRN), Tohoku University, Sendai 980-8577, Japan*

[6] *Graduate School of Environmental Studies, Tohoku University, Sendai 980-8579, Japan*

[7] *WPI-International Center for Materials Nanoarchitectonics (MANA), National Institute for Material Science, Tsukuba 305-0044, Japan*

[8] *School of Electrical and Computer Engineering and Birck Nanotechnology Center Purdue University, West Lafayette, Indiana 47907, USA*

[9] *Institute of Physics and Astronomy and Villum Center for Hybrid Quantum Materials and Devices, Aarhus University, 8000, Aarhus-C, Denmark*

[*]idzuchi@tohoku.ac.jp
#) H. Idzuchi and S. Iihama contributed equally to this work.

**Abstract**

Spin transport characteristics of graphene has been extensively studied so far. The spin transport along *c*-axis is however reported by rather limited number of papers. We have studied spin transport characteristics through graphene along *c*-axis with permalloy(Py)/graphene(Gr)/Pt by gigahertz (GHz) and terahertz (THz) magnetization dynamics driven by femtosecond laser pulses. The relatively simple sample structure does not require electrodes on the sample. The graphene layer was prepared by chemical vapor deposition and transferred on Pt film. The quality of graphene layer was characterized by Raman microscopy. Time resolved magneto-optical Kerr effect is used to characterize gigahertz magnetization dynamics. Magnetization precession is clearly observed both for Pt/Py and Pt/Gr/Py. The Gilbert damping constant of Pt/Py was 0.015, indicates spin pumping effect from Py to Pt. The Gilbert damping constant of Pt/Gr/Py is found to be 0.011, indicates spin injection is blocked by graphene layer. We have also performed the measurement of THz emission for Pt/Py and Pt/Gr/Py. While the THz emission is clearly observed for Pt/Py, a strong reduction of THz emission is observed for Pt/Gr/Py. With these two different experiments, and highly anisotropic resistivity of graphite, we conclude that the vertical spin transport is strongly suppressed by the graphene layer.

Recently, two-dimensional (2D) materials have attracted considerable attention. Two-dimensional materials provide handful access on highly crystalline samples, offering new spintronics research directions. Since spin currents can flow in nonmagnetic materials, so far such spin transport is widely studied in in-plane direction of nonmagnetic 2D materials [1,2,3]. In three-dimensional materials such as Pt, spin transport in out-of-plane direction is often studied with spin Hall effect. The spin current is converted to charge transport with certain geometry: the spin polarization, the flow direction of spin current and the direction of detection voltage need to be all perpendicular to each other. This geometrical constraint makes it difficult to study out-of-plane spin transport through *c*-axis of 2D material while the spin transport in-plane has been relatively well studied. Here, we optically investigated spin transport characteristics in *c*-axis of graphene by using gigahertz (GHz) and terahertz (THz) magnetization dynamics excited by a femtosecond pulse laser. This makes it more easily to satisfy such geometrical conditions as the injector and detector are not required to be electrically connected.

Figure 1 represents our sample structure as well as a brief measurement set up. Here, we employed spin pumping and THz emission, both induced by magnetization dynamics excited by a pulsed laser as described below. Previously, vertical spin transport in multilayers of graphene have been studied by ferro magnetic resonance. Patra et al fabricated the sample on a co-planer wave guide and used broad band frequency to characterize spin transport. They found the Gilbert damping is significantly enhanced for Py/Gr compared to Py/Pt where Py stands for permalloy ($Ni_{80}Fe_{20}$) [4]. Later Gannett et al studied series of the samples with different thickness of Py to characterize transport properties, which shows no detectable enhancement for Py/Gr/Cu [5]. While the interface of graphene can be complicated, in our experiment the sample structure is simple (just a multilayer film) and complimentary characteristics are obtained by two methods, which should help reveal the intrinsic interface spin transport properties.

In this study, spin transport was studied on Pt/graphene/Py and Pt/Py. Pt, graphene, and Py were chosen for representative materials for spin Hall effect, 2D material, and soft ferromagnet. Pt film was prepared by sputtering with the thickness of 3 nm on silicon substrate and glass substrate. The graphene film was transferred onto Pt in ambient condition. Graphene film was prepared on thin copper foil by a standard chemical vapor deposition (CVD) method and transferred onto the Pt film. Raman microscopy was used to characterize the number of layers in graphene where the laser wavelength is 532 nm. We have observed clear peaks of D, G, and 2D bands from left to right as shown in Fig.2a. Particularly from the 2D peak, we confirmed the crystallinity of the graphene film and the film was not folded [6,7]. The Py film and MgO capping layer was sputtered on graphene film with the base pressure of ~ $10^{-7}$ Torr. The static magnetization process of the film was examined by magneto optical Kerr effect. For measuring GHz magnetization dynamics induced by femtosecond laser pulse (Fig. 1a), time-resolved magneto-optical Kerr effect (TRMOKE) was employed [8]. The wavelength, pulse duration, and repetition rate for both the pump and probe laser pulses were 800 nm, 120 fs, and 1 kHz, respectively. The pump laser beam was irradiated on the sample from the film normal and the incident angle of the probe laser beam was ~ 5 degree measured from the film normal. Kerr rotation angle of the reflected probe beam was detected via balanced photo-detector. The pump laser pulse was modulated by the mechanical chopper with the frequency of 360 Hz and then the pump-laser induced change in Kerr rotation angle was detected by a Lock-in amplifier. A magnetic field was applied with a 10 degree angle measured from the film normal. The magnetization precession can be excited by the reduction of demagnetizing field due to laser heating. The damping of magnetization precession reflects the transfer of spin into adjacent normal metal layer, referred as spin-pumping effect [9, 10]. To study spin-transport induced by THz magnetization dynamics, THz time-domain spectroscopy was employed [11] (Fig. 1b), in which THz spin-current can be generated by ultrafast demagnetization of Py layer and its angular momentum can be transferred to Pt layer [12,13]. Then, THz electric field can be generated through spin-to-charge conversion (inverse spin Hall effect) in Pt layer. Wavelength, pulse duration, repetition rate for the laser pulse were 800 nm, 120 fs, and 80 MHz, respectively. The femtosecond laser was irradiated from substrate side and then THz electric field emitted from the film side was measured. The THz electric field was detected by electro-optic sampling method using a ZnTe (110) crystal. All measurements were conducted at room temperature.

The spin transport of graphene in vertical direction can be accessed by spin pumping with additional layers. We compare Pt/Py bilayer with Pt/graphene/Py trilayers to characterize spin transport properties across graphene layer. Figure 2b shows typical TRMOKE signal for Pt/Gr/Py(10nm)/MgO on Si substrate under the external magnetic field of 10.7 kOe in a direction tilted by 10 degrees from perpendicular to the substrate. We have clearly observed spin precession slowly decaying over long period right after initial ultrafast dynamics, for the samples of both with and without graphene layer. Those oscillations are fitted to the following equations

$$A + B \cdot \exp(-\nu t) + C \cdot \exp(-t/\tau) \sin(2\pi f t + \phi_0)$$

where, $A, B, \nu, C, f, \tau,$ and $\phi_0$ are signal offset, magnitude of exponential background signal due to recovery of magnetization, decay rate, oscillation amplitude, oscillation frequency, oscillation life-time, and initial phase, respectively. The TRMOKE signals are well fitted by the above equation shown as solid curve in Fig. 2(b). The $f$ and $1/\tau$ values evaluated by fitting with different applied magnetic fields are shown in Fig. 3. $f$ and $1/\tau$ can be calculated theoretically using Landau-Lifshitz Gilbert (LLG) equation as [8,14],

$$f_{\text{LLG}} = \frac{\gamma}{2\pi}\sqrt{H_1 H_2}, \tag{1}$$

$$\frac{1}{\tau_{\text{LLG}}} = \frac{1}{2}\alpha\gamma(H_1 + H_2), \tag{2}$$

$$H_1 = H\cos(\theta - \theta_H) - 4\pi M_{\text{eff}} \cos^2\theta, \tag{3}$$

$$H_2 = H\cos(\theta - \theta_H) - 4\pi M_{\text{eff}} \cos 2\theta, \tag{4}$$

where $H$, $\theta_H$ (=10 degree in this study), $\theta$, $4\pi M_{\text{eff}}$, $\gamma$, and $\alpha$ are external magnetic field, field angle, magnetization angle, effective demagnetizing field, gyromagnetic ratio and Gilbert damping constant respectively. $\gamma$ is given by the relation, $\gamma = g\mu_B/\hbar$. The $\theta$ is determined by the energy minimum condition as,

$$H \sin(\theta_H - \theta) + 2\pi M_{\text{eff}} \sin 2\theta = 0, \quad (5)$$

The measured $f$ is well fitted by Eq. (1) with the parameters $g$ = 2.09 (2.06) and $4\pi M_{\text{eff}}$ = -9.8 (-8.8) kOe for Pt / Gr / Py (Pt / Py) film. The $1/\tau$ calculated using Eq. (2) are shown in Fig. 3(b) and 3(c). $1/\tau$ for Pt / Gr / Py / MgO sample (Fig. 3(c)) can be well explained by Eq. (2) with $\alpha$ = 0.011. However, $1/\tau$ for Pt / Py / MgO cannot be explained by Eq. (2), which is due to inhomogeneous linewidth broadening. Therefore, $1/\tau$ enhancement due to inhomogeneous linewidth broadening is considered as follows,

$$\frac{1}{\tau_{\text{tot}}} = \frac{1}{\tau_{\text{LLG}}} + \frac{1}{2}\left|\frac{d(2\pi f_{\text{LLG}})}{d\theta_H}\right|\Delta\theta_H, \quad (6)$$

where, the first term is identical to Eq. (2) and the second term is $1/\tau$ enhancement due to distribution of $\theta_H$ which may be related to surface roughness of the film [14]. The black solid and blue dashed curves in Fig. 3(b) are the calculated results of the first and second terms of Eq. (6). $H_{\text{ext}}$ dependence of $1/\tau$ for Pt / Py / MgO filmis well explained by the summation of two contributions in Eq. (6) with the parameters, $\alpha$ = 0.015 and $\Delta\theta_H$ = 0.05 rad [green broken curve in Fig. 3(b)]. The enhancement of $\alpha$ is due to spin-pumping effect at Pt / Py interface associated with dissipation of angular momentum. This indicates strong suppression of spin current with graphene, consistent with Gannett et al [5]. Previously, it was shown that graphene has long spin diffusion length by means of lateral spin transport where spin current is flowing in-plane with long spin lifetime probed by Hanle effect [1,15]. Transport along $c$-direction can be rather different from the one in $ab$ plane. In early studies, graphite crystal shows rather anisotropic charge transport properties and the resistivity of $c$-axis is reported to larger than the one for $ab$ plane by a factor of $10^2$ to $10^3$ [16]. The resistive nature of the graphene along $c$ axis may prevent spin transport.

In the THz method, by irradiating femtosecond laser pulse on this kind of multilayers, THz electric field can be generated [12, 13]. Ultrafast spin current in nonmagnetic layer can be generated by the ultrafast demagnetization in Py layer, and spin-charge conversion via inverse spin-Hall effect in Pt layer create terahertz charge current and electric field. In our bilayer Pt/Py, we observed clear THz emission and its signal is inverted with reversed bias magnetic field, as shown in the top panel (a) of Fig. 4. Interestingly, on two Pt/graphene/Py samples, the THz signal was largely suppressed [Fig. 4(b) and 4(c)]. This implies strong suppression of spin injection from Py to Pt by graphene monolayer in the terahertz frequency region. The interpretation is qualitatively consistent with spin pumping study (using ultrafast laser heating and GHz magnetization dynamics). The strong reduction of THz signal is attributed to the strong suppression of spin-transport by inserting graphene monolayer with high resistivity along the $c$-axis. Strong reduction of THz emission was also reported for Co/ZnO/Pt junctions[17]. With these two different characterization methods, we conclude graphene monolayer effectively blocks vertical spin current. For the second sample of Pt/graphene/Py (Fig.4c), a small peak appeared around 1 ps. We notice this may or may not be a delayed THz emission, whose precise mechanism (e.g., how it may be related to the graphene barrier, whether it may also be generated by Py itself etc.) is not clear yet at this stage and open for future study.

In conclusion, we have investigated spin injection characteristics of Py/graphene/Pt by means of gigahertz and terahertz magnetization dynamics driven by a femtosecond laser pulse. Graphene layer was grown by CVD method and the Raman characteristics on Pt showed the characteristics of single layer graphene film. We have clearly observed GHz magnetization precession induced by the laser pulse for the samples of both with and without graphene (Py/Pt). Graphene is observed to give an apparent suppression of the damping enhancement due to spin-pumping effect at Py / Pt interface, indicating reduction of angular momentum dissipation by graphene monolayer. THz emission induced by femto-second laser pulse was observed for Py/Pt bilayer, while the THz emission was strongly suppressed for Py/graphene/Pt, which clear indicates graphene blocks spin current in transport along $c$-axis. Both experiments on spin pumping and THz method can be understood by the strongly suppressed spin transport across the graphene layer.

**Data Availability Statements**

The data that support the findings of this study are available from the corresponding author upon reasonable request.

**Acknowledgments**

We acknowledge the support from AIMR common equipment unit. This work was supported in part by Advanced Institute for Materials Research (AIMR) under World Premier International Research Center Initiative (WPI) of MEXT, Japan, and by AIMR fusion research program, by the Mazda Foundation, and by a Grant-in-Aid for Scientific Research from the Ministry of Education, Culture, Sports, Science and Technology (MEXT), JSPS KAKENHI (Grant Number 18H03858, 18H04473, 20H04623, and 20K14399).

**Figures**

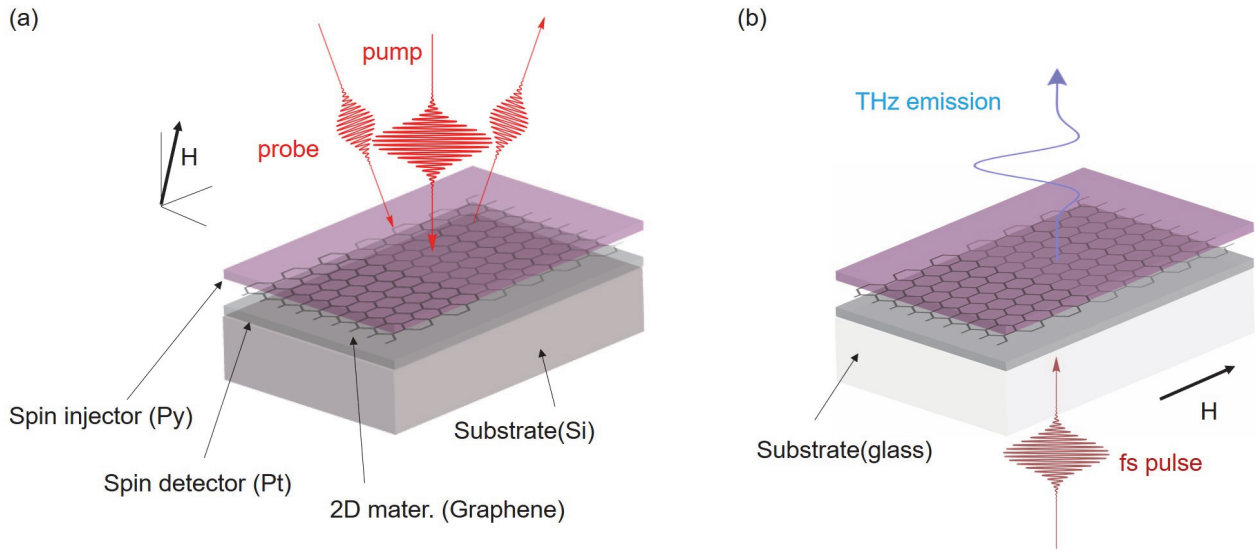

**Fig. 1.** Concept and schematic image of experimental set up of this study. Graphene, spin injector (Py) and detector (Pt) are depicted in black hexagon, purple box, and gray box, respectively. (a) The set up for pump-probe measurement for magnetization dynamics. The probe beam is slightly (~ 5 deg) tilted from the film normal. The magnetic field is applied with a 10 degree angle measured from the film normal. (b) The set up for THz emission. The magnetic field is applied in plane.

Fig.1

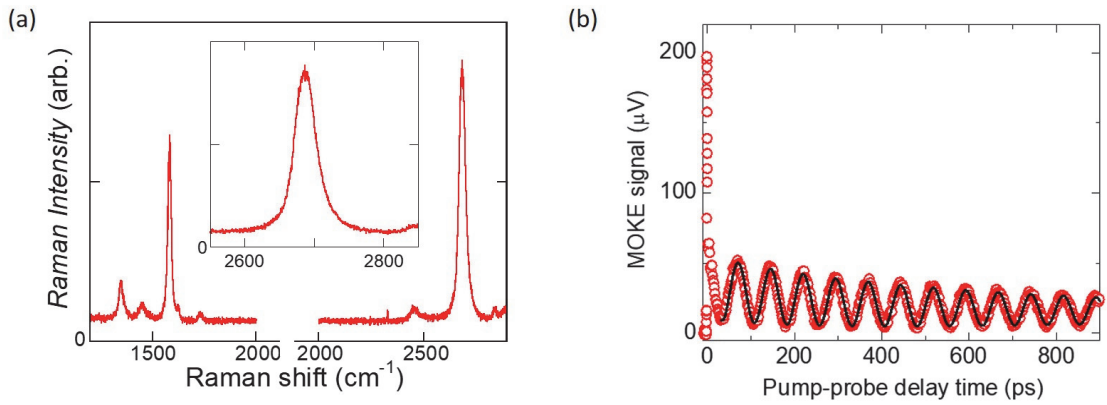

**Fig. 2.** (a) Raman spectroscopy of a graphene layer transferred on to a glass/Pt substrate (glass with Pt sputtered). Three main Raman peaks (D, G, and 2D) are clearly observed. Inset shows the 2D peak, clearly different from bilayer or multilayers graphene [6]. (b) Magneto-optical Kerr effect (MOKE) signal plotted as a function of pump-probe delay time for Pt/Gr/Py(10nm)/MgO under an external magnetic field of 10.7 kOe. The field is applied with a 10degree angle measured from the film normal. The measurement set up is schematically shown in Fig.1a.

Fig.2

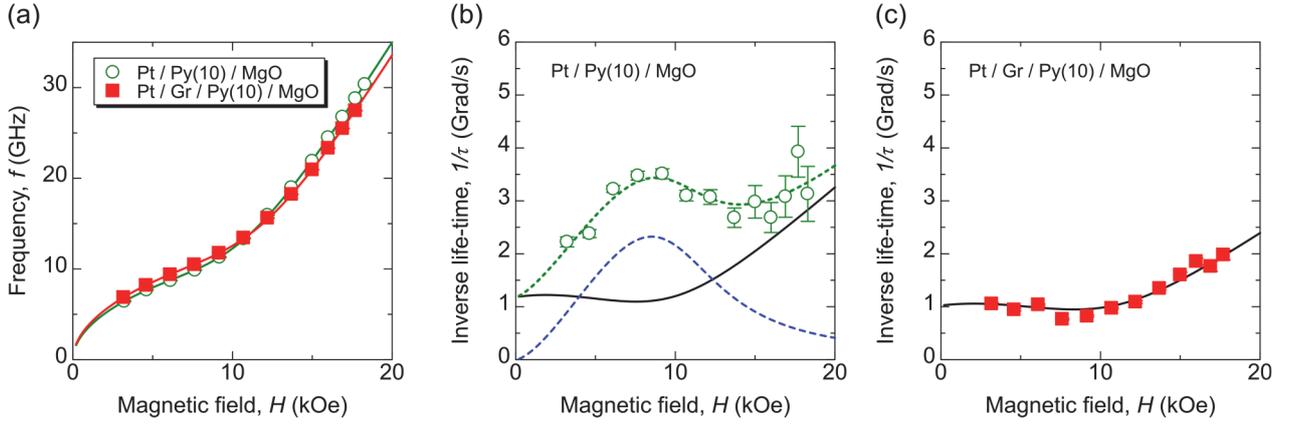

**Fig. 3.** Characterization of GHz magnetization dynamics for multilayers with and without graphene. The measurement set up is schematically shown in Fig.1a. (a) precession frequency as a function of the magnetic field. The field is applied with a 10 degree angle measured from the film normal. Closed red squares and open green circles indicate data from the Pt/Gr/Py/MgO and Pt/Py/MgO respectively where the thickness of Py is 10 nm for both case. The solid curves are obtained by Eq. (1) with the parameters in the main text. Inverse lifetime as a function of the magnetic field for (b) Pt / Py / MgO and (c) Pt / Gr / Py / MgO films. The black solid curves shown in (b) and (c) correspond to the calculated value using LLG eq. (Eq. (2)). The green dotted and blue broken curves shown in (b) are the left hand side and the second term in the right hand side in Eq. (6), respectively, with the parameters in the main text.

Fig.3

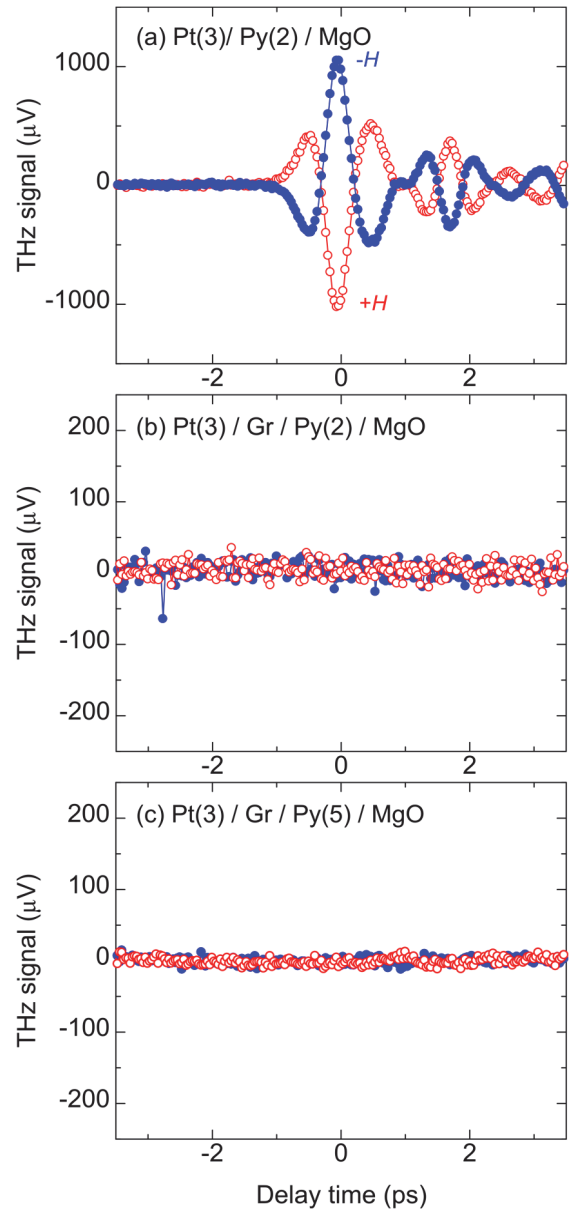

**Fig. 4.** Detection of THz electric field generated by femto-second laser pulse on (a) Pt(3) / Py(2) /MgO , (b) Pt(3) / Gr / Py(2) / MgO and (c) Pt(3) / Gr / Py(5) / MgO made on glass substrates. The measurement set up is schematically shown in Fig.1b. The numbers in bracket indicate the thickness of the layers in the unit of nanometers. The magnetic field was applied in in-plane direction. Blue solid and red open symbols are the signal obtained with opposite magnetic field direction.

Fig.4

11